\documentclass[sigconf]{acmart}

\AtBeginDocument{%
  }

\setcopyright{acmlicensed}
\copyrightyear{2018}
\acmYear{2018}
\acmDOI{XXXXXXX.XXXXXXX}
\acmConference[Conference acronym 'XX]{Make sure to enter the correct
  conference title from your rights confirmation email}{June 03--05,
  2018}{Woodstock, NY}
  \setcopyright{acmcopyright}

\acmISBN{978-1-4503-XXXX-X/2018/06}

\begin{document}
\title{Positioning Modular Co-Design in Future HRI Design Research}

\author{Lingyun Chen}
\affiliation{
  \institution{Luddy School of Informatics, Computing, and Engineering, Indiana University Bloomington}
  \city{Bloomington}
  \state{IN}
  \country{USA}
}
  \email{lch2@iu.edu}

\author{Qing Xiao}
\affiliation{
  \institution{Human-Computer Interaction Institute, Carnegie Mellon University}
  \city{Pittsburgh}
  \state{PA}
  \country{USA}
}
  \email{qingx@andrew.cmu.edu}

\author{Zitao Zhang}
\affiliation{
  \institution{Luddy School of Informatics, Computing, and Engineering, Indiana University Bloomington}
  \city{Bloomington}
  \state{IN}
  \country{USA}
}
  \email{zhangzit@iu.edu}

\author{Eli Blevis}
\affiliation{
  \institution{Luddy School of Informatics, Computing, and Engineering, Indiana University Bloomington}
  \city{Bloomington}
  \state{IN}
  \country{USA}
}
  \email{eblevis@iu.edu}

\author{Selma Šabanović}
\affiliation{
  \institution{Luddy School of Informatics, Computing, and Engineering, Indiana University Bloomington}
  \city{Bloomington}
  \state{IN}
  \country{USA}
}
  \email{selmas@iu.edu}

\renewcommand{\shortauthors}{Chen et al.}

\begin{abstract}
Design-oriented HRI is increasingly interested in robots as long-term companions, yet many designs still assume a fixed form and a stable set of functions. We present an ongoing design research program that treats modularity as a designerly medium—a way to make long-term human–robot relationships discussable and material through co-design. Across a series of lifespan-oriented co-design activities, participants repeatedly reconfigured the same robot for different life stages, using modular parts to express changing needs, values, and roles. From these outcomes, we articulate PAS (Personalization–Adaptability–Sustainability) as a human-centered lens on how people enact modularity in practice: configuring for self-expression, adapting across transitions, and sustaining robots through repair, reuse, and continuity. We then sketch next steps toward a fabrication-aware, community-extensible modular platform and propose evaluation criteria for designerly HRI work that prioritize expressive adequacy, lifespan plausibility, repairability-in-use, and responsible stewardship—not only usability or performance.
\end{abstract}



\maketitle

\section{Introduction}
Robots are increasingly imagined as long-term companions that accompany people across everyday life rather than short-lived tools optimized for isolated tasks \cite{leite2013social, kaplan2005everyday,chen20253r, bemelmans2012socially}. Yet much of HRI still evaluates robots as relatively stable products: designed for a fixed context, a fixed role, and a bounded timeframe \cite{carstensen1999taking, fung2001age}. This creates a fundamental mismatch \cite{irfan2023lifelong, kidd2008robots}. Human lives are dynamic. Our capabilities, responsibilities, environments, and relationships shift across the lifespan \cite{carstensen1999taking, fung2001age}. While most robots remain closed and difficult to evolve \cite{vsabanovic2010robots, koike2024sprout,tang2025robolinker}. As a result, robots that once "fit'' often become misaligned, underused, or discarded as needs change, contributing to premature obsolescence and e-waste \cite{leite2013social, hansson2021decade, remy2014addressing, yu2024if}.

We argue that enabling \textit{lifespan HRI} requires expanding what we mean by sustainability in HRI beyond energy efficiency or durable hardware \cite{vsabanovic2010robots, ye2025game}. Long-term HRI depends on both \textit{material continuity} (a robot can be maintained \cite{mcgloin2024consulting}, repaired, and upgraded without full replacement) and \textit{relational continuity} (a robot can remain meaningful as identities, roles, and contexts change) \cite{ye2025game}. Importantly, these two forms of continuity are intertwined: people are more willing to repair and keep artifacts that still feel like "the same'' companion and still fit their lives.

This position paper advances the claim that modular co-design is a practical pathway to robots that can evolve with us. We treat modularity not merely as an engineering strategy but as a \textit{socio-technical practice} through which users can reconfigure form and function over time \cite{alattas2019evolutionary}, shifting a robot's role, preserving identity-critical parts, and engaging in repair and reuse as everyday interaction \cite{alattas2019evolutionary, article}. To ground this argument, we build on empirical co-design workshops with younger and older adults (N=23) in which participants used modular building blocks to imagine robots across life stages. Their designs consistently challenged fixed-role assumptions and revealed how users conceptualize continuity, change, and care through modular reconfiguration.

Our contribution is threefold. First, we propose PAS (Personalization, Adaptability, Sustainability) as a human-centered framing that explains how modularity becomes meaningful in lived HRI, and how it can support both material and emotional durability over time. Second, we position modular co-design as a temporal probe—a method for eliciting how people anticipate role shifts and negotiate identity and attachment across life transitions. Third, we outline implications and provocations for the HRI community, including a shift from product thinking to platform thinking, evaluation criteria that emphasize long-term relationship resilience, and sustainability treated as a core interaction outcome.

The remainder of this paper introduces PAS, summarizes empirical insights from lifespan modular co-design, and offers workshop-oriented discussion prompts aimed at researchers and designers working on sustainable long-term human--robot relationships.

This raises a central challenge for Human–Robot Interaction (HRI): how can robots remain relevant, meaningful, and sustainable as peoples’ lives advance?

\section{From SAM to PAS: A Human-Centered Framework}
Prior frameworks such as SAM (Sustainability–Adaptability–Modularity) emphasize the link between modularity and sustainability, but often remain at a speculative or designer-centric level. Building on empirical co-design insights, we propose PAS as a more directly human-centered articulation of how modularity becomes meaningful in lived HRI.

\subsection{PAS Framework}
\paragraph{Personalization}
personalization is not cosmetic customization, but a human-centered process through which people use modular configurations to inscribe roles, preferences, and values into a robot and continually adjust it to situated life circumstances. Crucially, our data suggests personalization provides the direction and motivation for long-term use: when reconfigurations express identity, roles, and values, they cultivate \textit{“for me / part of my life / worth keeping”} meaning-making, which increases ownership and attachment and makes people more willing to maintain, repair, and carry modules forward rather than abandon the robot. 

\paragraph{Adaptability}
PAS treats adaptability as a robot’s capacity to evolve alongside users’ shifting social, emotional, and functional needs, enabled through people’s reconfiguration of form and function as life events and responsibilities shift. 

Empirically, adaptability appears in two temporal scales:
1. Within-stage adaptability: flexibly switching roles to meet overlapping demands in the same life stage (e.g., balancing productivity and care in adulthood). 
2. Lifespan evolution: sustaining one evolving robot across decades through two modes—radical reinvention (complete shifts across stages) and selective continuity (retaining key modules to preserve familiarity/attachment while swapping others).

\paragraph{Sustainability}
sustainability is not only about “reducing waste,” but about keeping robots relevant, meaningful, and ecologically responsible over the long term, with modularity preventing obsolescence by allowing parts to be swapped, reused, and reinterpreted. 

Participants articulated sustainability as two intertwined dimensions:
1. Material sustainability: repair, reuse, evolve/upgrade modules without discarding the whole system—normalizing repair and reuse as everyday practices. 
2. Emotional sustainability: continuity, attachment, identity—keeping familiar modules over time to preserve shared history and “same robot” recognition, which further supports willingness to maintain and repair.

\begin{figure}
    \centering
    \includegraphics[width=1\linewidth]{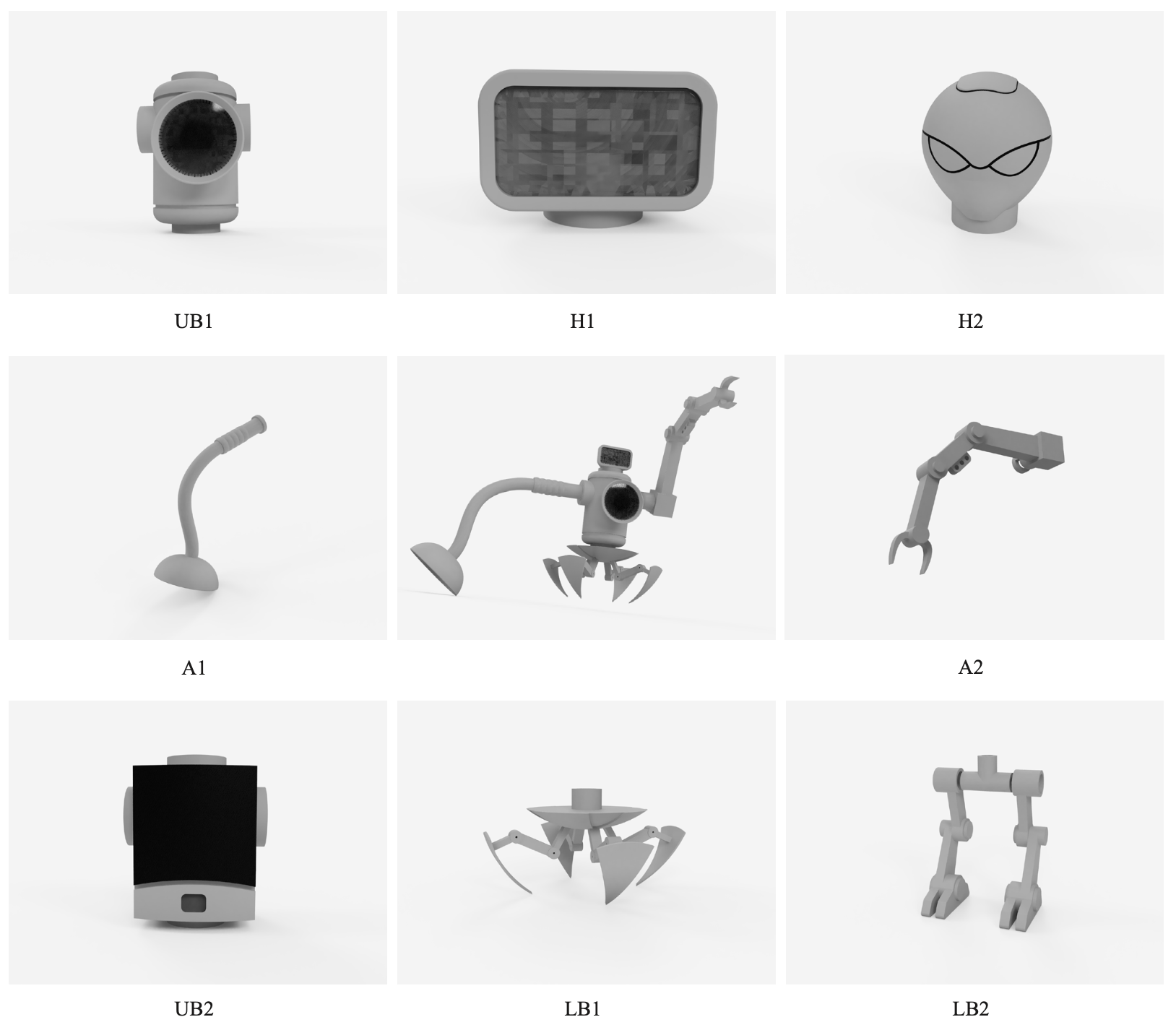}
    \caption{A list of selectively 3D appearance modules, translated from a collection of digital sketches, illustrate how our proposed parts can be assembled into different robot configurations.}
    \label{fig:collection}
\end{figure}

\section{Empirical Insights from Lifespan Modular Co-Design}
We conducted co-design workshops with 15 younger adults and 8 older adults (N=23). When given modular building blocks and prompts to imagine robots across life stages, participants produced concepts that exceeded typical fixed-role HRI assumptions. To make lifespan reconfiguration concrete, we worked with a simple, body-like modular scaffold with four part families—Head (H), Upper Body (UB), Arms (A), and Lower Body (LB)—as a starting assumption rather than an end goal (see \autoref{fig:collection}. Participants could swap, omit, and recombine these parts across life stages, and were also encouraged to break the structure when it did not fit their intentions.

\subsection{Role Fluidity Across Life Stages}
Across the workshops, participants rarely imagined a robot that should stay in one “final” form. Instead, they described a persistent core that can be re-authored through replaceable modules as life priorities shift. In childhood, designs leaned into play and protection. One participant described a robot as \textit{“a doll that can jump, play music… Tablet head to play cartoon, soft arm to hug, body for protection, leg to play in parks,”} explicitly tying modules to what childhood companionship should feel like. 

In adulthood, the same “robot companion” was expected to support efficiency, work, and daily load. A participant framed this stage as needing \textit{“something that helps me in my career and fasten my daily tasks,”} signaling a turn toward productivity-oriented configurations rather than playful ones. 

In older adulthood, participants shifted again toward care, safety, and mobility support. For example, one participant emphasized long-term assistance through a practical carrying/storage function:\textit{ “A box to hold things like medicine, water, and food… help me carry items, like a shopping cart.” }Another described older-adult configurations as spanning \textit{“hearing… socialization… watch for falling,”} combining assistance and connection rather than a single-purpose “eldercare robot.” 

\subsection{Attachment Through Continuity}
Even when most modules were replaced, participants often insisted the robot could remain “the same” companion if a meaningful core persisted (e.g., a distinctive face module, memory module, or a symbolic piece retained across time). This continuity was central to emotional attachment: identity became relational and narrative, not strictly material. Some participants framed modular robots as potential heirlooms—objects with intergenerational continuity rather than disposable electronics. One participant described how keeping familiar modules creates recognition and memory:\textit{ “you still have some old pieces or old modules that you will recognize from before—nostalgia.” }

They also made the continuity argument explicitly through family/toy metaphors. A participant said, \textit{“Sometimes I still keep the toy. So if you change the robot, it’s just changed part of it, like keeping a family member,”} framing identity as something that persists even when components change. Finally, some participants linked continuity to self-identity, describing modular change as evolution rather than replacement. One participant explained:\textit{ “It felt like it’s an extension of me… having the ability to move the pieces would feel like… the same me, but it’s just evolving… as opposed to… buying a new phone every time and I’m not attached.” }

This directly supports that attachment can be maintained through a stable “core” (face, memory, signature element) even as other modules rotate over time.

\subsection{Repair as an Interaction Ritual}
Participants repeatedly framed repair and replacement not as a frustrating exception, but as a normal and valuable way of living with technology—especially when the system is designed for easy swapping. One participant criticized the throwaway norm and positioned modular exchange as a better relationship with devices: \textit{“Having parts to exchange is always better than throwing the whole thing away… If something breaks, you basically just throw it away.”} Repair was described as a way to preserve relationship continuity, not just function. A participant said: \textit{“I would like to repair a current robot and keep continuity and expectations. It would grow old with me, like a friend.”} Participants reframed repair not as breakdown management, but as a \textit{relationship practice}. When replacement is simple and modular, fixing becomes an act of care that strengthens ownership and attachment. In this sense, repair is not outside HRI—it is a form of HRI.

\section{Implications and Provocations for the HRI Community}

We propose three shifts for the HRI community to enable sustainable lifespan HRI:

\subsection{From Product Thinking to Platform Thinking}
If robots are meant to stay in people's lives for years or decades, then designing them as closed, finished products is a mismatch. A platform-oriented view treats the robot as an evolving ecosystem: a stable core plus swappable modules, expandable capabilities, and a community of contributors. This enables two forms of growth: users can adapt the robot to their changing lives. Designers and developers (including third parties) can extend the robot beyond what the original team anticipated.

Moving to platforms raises practical questions. What should count as a module—physical parts, behaviors, or both? What basic standards (connectors, power, data) enable swapping modules without locking users into one company? How can we keep systems compatible while still allowing diverse forms and interaction styles? Finally, we should consider participation: how do we credit community contributions and make modular creation accessible beyond technically confident users?

A platform-oriented research agenda suggests new workshop directions: design patterns for modular robot ecosystems; “minimum viable standards” for interoperability; and participatory governance models that balance openness with safety, accountability, and equity.

\subsection{Designing for Change}
Lifespan robots will be reconfigured, repaired, and upgraded over time. These changes are not exceptions—they are normal. Designing for change means expecting robots to shift with everyday realities, such as household routines, caregiving needs, changing abilities, and available parts.

This has two design implications. First, reconfiguration should be an easy, supported interaction, not a hidden maintenance task. Users should be able to understand what a module does, what it affects, and how to undo changes if something goes wrong. Second, systems should support continuity: even if parts change, people may want the robot to still feel like the same companion through stable identity cues (e.g., voice, face, memories, routines).

Evaluation should also move beyond short-term task success. We need ways to study how robots hold up through changes—after upgrades, breakdowns, or role shifts—and whether users can adapt them without losing trust or attachment. Workshop discussion can focus on practical methods for evaluating transitions and design patterns for safe, user-driven evolution.

\subsection{Sustainability as a Core Interaction Metric}
Sustainability is often treated as an external constraint—materials, manufacturing, energy use—rather than something interaction design can directly influence \cite{blevis2007sustainable, disalvo2010mapping}. A lifespan perspective argues the opposite: interaction design strongly shapes whether a robot is maintained, repaired, repurposed, or discarded. Sustainability should therefore be treated as an interaction outcome, alongside usability, safety, and user experience.

This framing shifts attention to behaviors and decision points across the robot’s life: what happens when a component fails, when needs change, when a newer model appears, when software support ends, or when the robot no longer fits a household routine. Design choices can either accelerate replacement (sealed designs, opaque diagnostics, forced upgrades) or support longevity (repairable parts, transparent health status, modular upgrades, easy reconfiguration, and meaningful “keep” incentives such as preserved memories or identity continuity).

Treating sustainability as a metric invites new measures the HRI community can develop and standardize. Beyond energy consumption, we can evaluate longevity (time-in-use, retention across life transitions), repairability (time-to-repair, accessibility of repairs, success rates), reuse and upgrade patterns (module carry-over rates, recombination frequency), and end-of-life outcomes (resale, donation, parts harvesting, responsible recycling). Importantly, sustainability metrics should include human factors: effort, frustration, confidence, and perceived value—because people do not repair what feels impossible, humiliating, or not worth it.

\section{Conclusion}
We position modular co-design as a designerly pathway toward robots that evolve with people: personalized, adaptable, and sustainable in both material and emotional senses. This position paper contributes a workshop-ready framing (PAS), a method (modular co-design as temporal probe), and a set of evaluation merits intended to support constructive discussion among HRI design researchers working on long-term human--robot relationships.

\bibliographystyle{ACM-Reference-Format}
\bibliography{references}

\end{document}